\documentclass{revtex4}

\usepackage{graphicx}
\begin{document}
% You should use BibTeX and revtex.bst for references
\bibliographystyle{revtex}

% Use the \preprint command to place your local institutional report
% number  and your conference paper identification number on the
% title page in preprint mode. Multiple \preprint commands are allowed.
\preprint{CERN-EP/2001-XX}

%Title of paper
\title{Studying the Higgs Potential at the $e^+e^-$ Linear Collider}
%\title[]{}

% repeat the \author .. \affiliation  etc. as needed
% \email, \thanks, \homepage, \altaffiliation all apply to the current
% author. Explanatory text should go in the []'s, actual e-mail
% address or url should go in the {}'s for \email and \homepage.
% Please use the appropriate macro for the type of information

% \affiliation command applies to all authors since the last
% \affiliation command. The \affiliation command should follow the
% other information

\author{Marco Battaglia}
\email[]{Marco.Battaglia@cern.ch}
\affiliation{CERN, CH-1211 Geneva 23, Switzerland}
\author{Eduard Boos}
\email[]{boos@theory.sinp.msu.ru}
\affiliation{Institute of Nuclear Physics, Moscow State University, 119899 Moscow, 
Russia}
\author{Weiming Yao}
\email[]{wmyao@lbl.gov}
\affiliation{LBNL, Berkeley CA 94720, USA}

%\collaboration{}
%\noaffiliation

\date{October 15, 2001}

\begin{abstract}
\vspace*{0.25cm}
The determination of the shape of the Higgs potential is needed to complete the 
investigation of the Higgs profile and to obtain a direct experimental proof of the 
mechanism of electro-weak symmetry breaking. This can be achieved, at a linear collider,
by determining the Higgs triple self-coupling $g_{HHH}$ in the processes  
$e^+e^- \rightarrow H^0H^0Z^0$ and $H^0H^0\nu_e \bar \nu_e$ and, possibly, the quartic 
coupling. This paper summarises the 
results of a study of the expected accuracies on the determination of $g_{HHH}$ at a 
TeV-class LC and at a multi-TeV LC. The statistical dilution arising from contributions
not sensitive to the triple Higgs vertex, can be reduced by means of variables sensitive
to the kinematics and the spin properties of the reactions.

\end{abstract}
% \pacs{}

%\maketitle must follow title, authors, abstract and \pacs
\maketitle

% References should be done using the \cite, \ref, and \label commands
\section{Introduction}

The detailed study of the Higgs potential represents a conclusive test of the 
mechanism of symmetry breaking and mass generation. After the discovery
of an elementary Higgs boson and the test of its couplings to quarks, leptons and 
gauge bosons, a further proof of the Higgs mechanism will be the experimental evidence 
that the Higgs field potential has the properties required for breaking the electro-weak
symmetry. In the Standard Model (SM), the Higgs potential can be written as
$V(\Phi^* \Phi)={\lambda} (\Phi^*\Phi - \frac{1}{2}v^2)^2$.
A probe of the shape of this potential comes from the determination of 
the triple and quartic Higgs self-couplings~\cite{Boudjema:1996cb,Ilin:1996iy,
Djouadi:1999gv,Djouadi:1999ei}.
The triple coupling can be expressed, in the SM, as 
$g_{HHH} = \frac{3}{2} \frac{M_H^2}{v}$ where $v$=246~GeV and $M_H$ can be determined 
to ${\cal{O}}(100~{\mathrm{MeV}})$ accuracy. An accurate test of this relation may
reveal the extended nature of the Higgs sector. This can be achieved by observing the 
deviations, arising in a generic 2HDM or in a SUSY scenario, from the SM prediction 
above.  Accurate data can be analysed in terms of an effective Lagrangian to establish 
the relationships between the Higgs self-couplings and the size of anomalous terms of 
other nature.

This paper summarises the findings of a comparative study of the potential of a 
TeV-class $e^+e^-$ linear collider (LC) and of a second-generation multi-TeV LC in
the study of the Higgs potential through the analysis of the  
$e^+e^- \rightarrow H^0H^0Z^0$ and $H^0H^0\nu_e \bar \nu_e$ processes in SM.
The opportunity, offered by a LC, to study the Higgs potential may be unique, as no 
evidence that the SM triple Higgs vertex is experimentally accessible at hadron 
colliders has been obtained so far~\cite{Lafaye:2000ec}. 

At the linear collider the triple Higgs coupling $g_{HHH}$ can be accessed by studying 
multiple Higgs production in the processes $e^+e^- \rightarrow H^0H^0Z^0$ and 
$H^0H^0\nu_e \bar \nu_e$ that are sensitive to the triple Higgs vertex. The first 
process is more important at lower values of the centre-of-mass energy, $\sqrt{s}$, and 
for Higgs boson masses in the range 115~GeV~$< M_H <$~130~GeV, with a cross section 
of the order of 0.15~fb for $M_H$ = 120~GeV. 
The second process becomes sizeable at collision energies 
above 1~TeV and ensures sensitivity to the triple Higgs vertex for heavier Higgs masses. 
It also provides a cross section larger by a factor $\simeq 7$. 
The quartic Higgs coupling remains elusive. The cross sections for the $HHHZ$ and 
$HHH\nu\bar{\nu}$ processes are reduced by three order of
magnitude compared to those for the double Higgs production. In the most favourable 
configuration, a 10~TeV $e^+e^-$ collider operating with a luminosity of 
$10^{35}$cm$^{-2}$s$^{-1}$ would be able to produce only about five such events in one 
year (=10$^7$s) of operation for $M_H$ = 120~GeV (see Table~\ref{tab:hhhnn}).

A $\gamma \gamma$ collider has also access to the triple Higgs couplings through the 
processes $\gamma \gamma \rightarrow HH$ and $\gamma \gamma \rightarrow HHW^+W^-$. 
However the cross section of the first process is suppressed by the effective 
$H \gamma \gamma$ coupling, compared to $e^+e^-$ collisions, and that of the second is 
only 0.35~fb at $\sqrt{s_{ee}}$ = 1~TeV for $M_H$ = 120~GeV. A muon collider does not 
offer significant advantages, compared to a high energy $e^+e^-$ LC. The most favourable 
production cross sections are comparable: the 
$\mu^+\mu^- \rightarrow HH\nu_{\mu}\bar{\nu_{\mu}}$ process has  
$\sigma$=0.8~fb-2.7~fb for $\sqrt{s}$ = 3~TeV-7~TeV. On the contrary, the 
$\mu^+\mu^- \rightarrow HH$ reaction, which would be unique to a muon collider, is 
strongly suppressed. 

A major problem arising in the extraction of $g_{HHH}$ from the measurement of the 
double Higgs production cross section comes from the existence of diagrams that lead to 
the same final states but do not include a triple Higgs vertex. The resulting dilution 
$\frac{\delta g_{HHH}/g^{SM}_{HHH}}{\delta \sigma/\sigma^{SM}}$, is significant for all 
the processes considered. It is therefore 
interesting to explore means to enhance the experimental sensitivity to the signal 
diagrams with additional variables. In this study the $H$ decay angle in the $HH$ rest 
frame and the $HH$ invariant mass have been considered.

The cross section computation and the event generation at parton level have been 
performed using the {\sc CompHEP}~\cite{Pukhov:1999gg} 
program where the $g_{HHH}$ coupling 
and the $M_H$ mass have been varied. Partons have been subsequently fragmented 
according to the parton shower model using {\sc Pythia}~6~\cite{Sjostrand:2001yu}. 
The resulting final state particles, for the $HH\nu\bar{\nu}$ channel, have been 
processed through the {\sc Simdet}~\cite{Pohl:1999uc} 
parametrised detector simulation and the reconstruction analysis performed. 
Background processes for 
the $HH\nu\bar{\nu}$ channel have also been simulated and accounted in the analysis.
The centre-of-mass energies of 0.5~TeV / 0.8~TeV and 3~TeV, representative of the 
anticipated parameters for a TeV-class LC, such as {\sc Tesla} or a X-band collider,
and a multi-TeV collider, such as {\sc Clic}, have been 
chosen with an integrated luminosity of 1~ab$^{-1}$ and 5~ab$^{-1}$ respectively. 
In this study, unpolarised beams are assumed. Polarizing the beams brings a gain of a
factor of two for the $HHZ$ cross section and of four for $HH\nu\bar{\nu}$.

\section{The $HHZ$ process}

\begin{figure}[t]
\begin{tabular}{c c}
\includegraphics[scale=0.4]{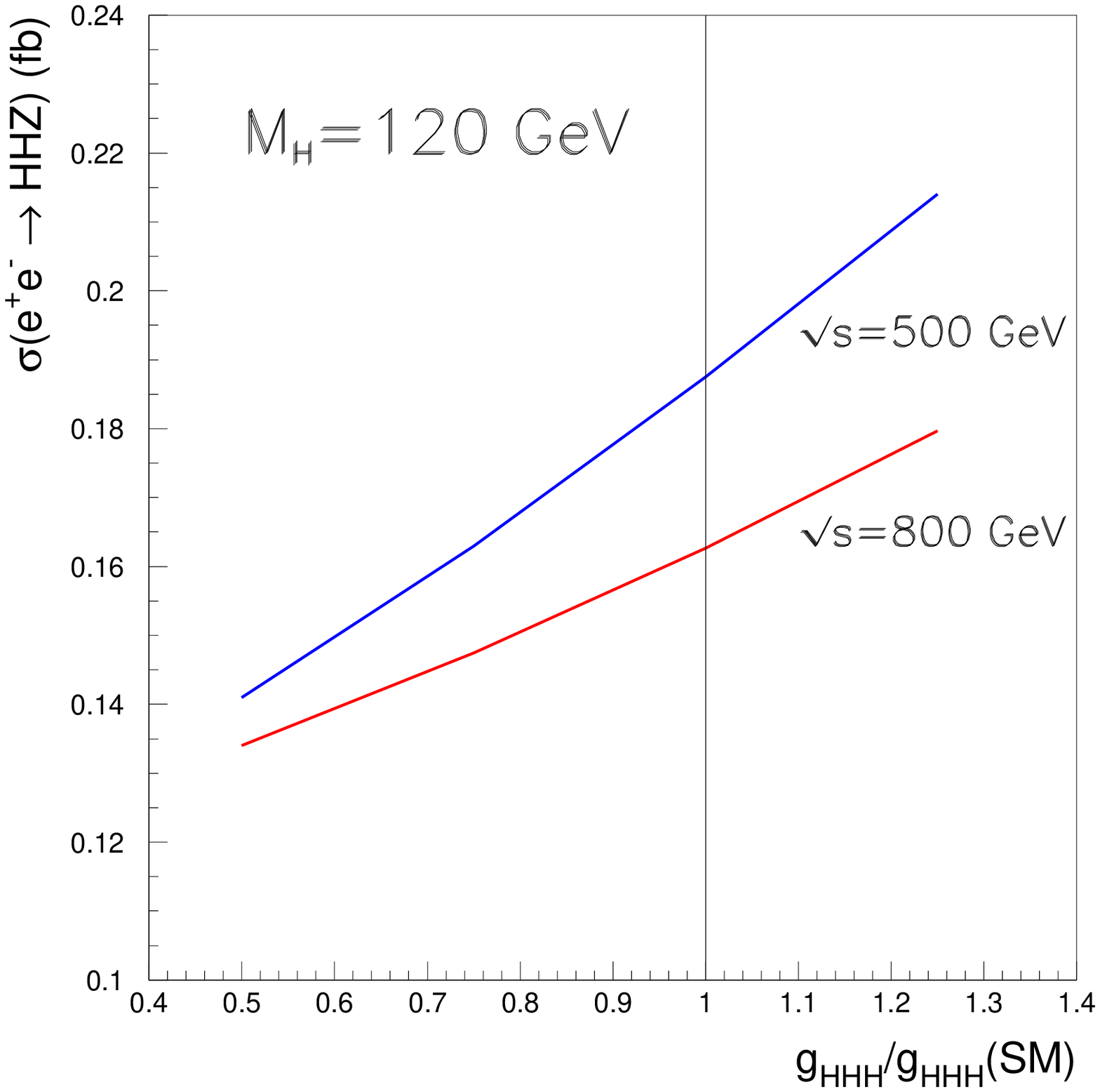} &
\includegraphics[scale=0.4]{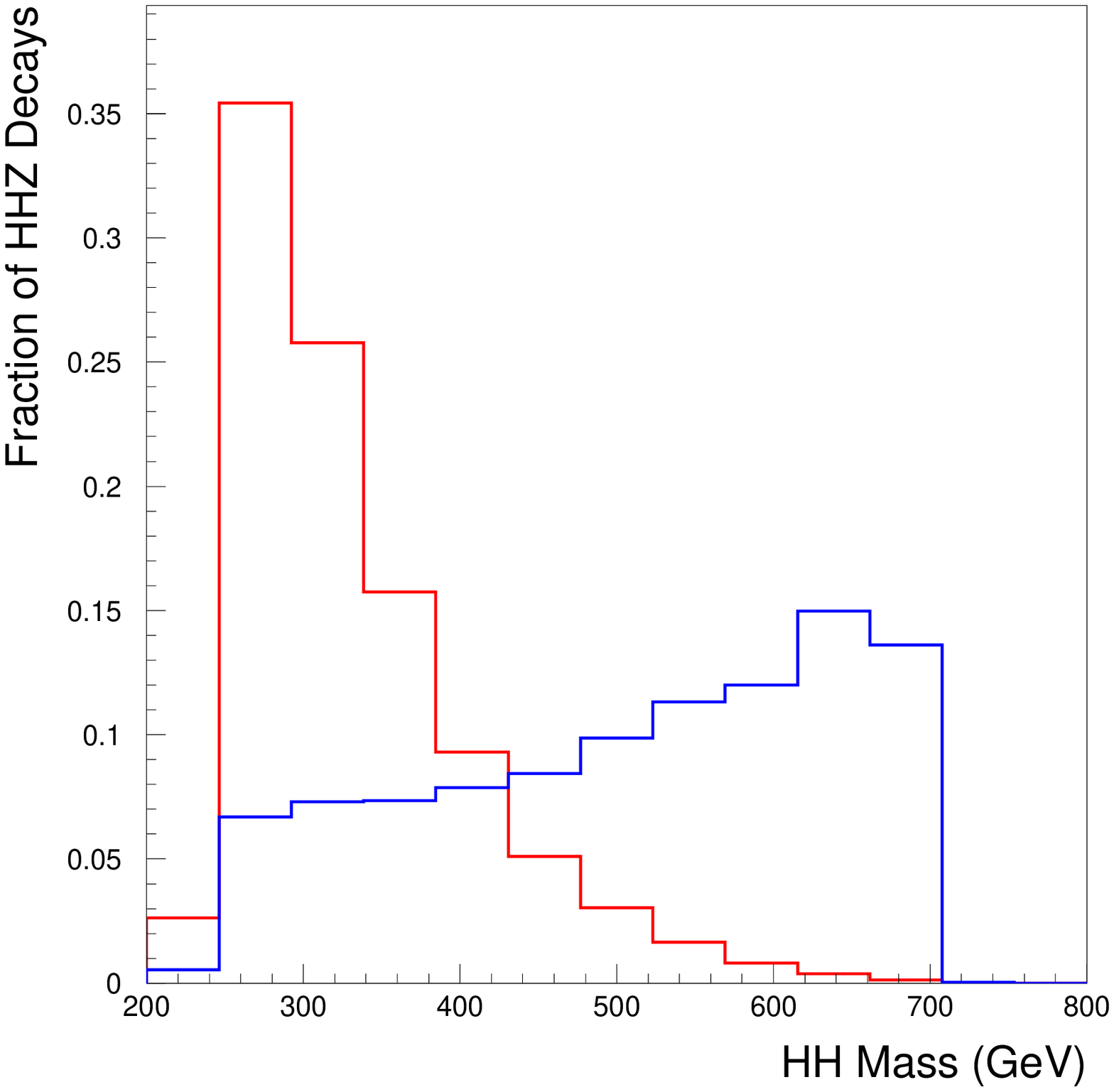} \\
\end{tabular}
\caption{\sl Left: the dependence of the $HHZ$ cross section on the triple Higgs 
coupling, normalised to its SM value, for $M_H$ = 120~GeV and two $\sqrt{s}$ values. 
Right: The distribution of the $HH$ invariant mass in $HHZ$ events orginating from 
diagrams containing the triple Higgs vertex (light grey) and other diagrams (dark grey).}
\label{fig:hhz1}
\end{figure}

\begin{figure}[b]
\includegraphics[scale=0.5]{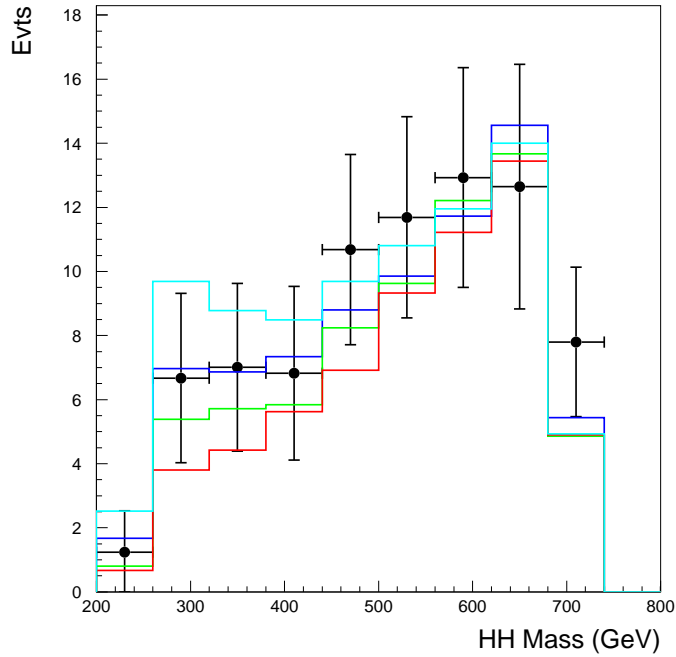}
\caption{The generated $HH$ invariant mass distribution for $HHZ$ events obtained for 
$g_{HHH}/g_{HHH}^{SM}=$1.25,1.0,0.75 and 0.5 with the points with error bars showing 
the expectation for 1~ab$^{-1}$ of SM data at $\sqrt{s}$=0.8~TeV.}
\label{fig:hhz2}
\end{figure}

The $g_{HHH}$ determination from the measurement of the $e^+e^- \rightarrow HHZ$ 
cross-section has been already extensively discussed in its 
theoretical~\cite{Djouadi:1999gv} 
and experimental issues~\cite{Castanier:2001sf}. 
A $e^+e^-$ collider operating at $\sqrt{s}$ = 
500~GeV can measure the $HHZ$ production cross section to about 15\% accuracy if the 
Higgs boson mass is 120~GeV, corresponding to a fractional accuracy of 23\% on $g_{HHH}$.
At higher $\sqrt{s}$ energies the dilution increases and it becomes interesting to 
combine a discriminating variable. 
The invariant mass of the $HH$ system provides with a 
significant discrimination of the $H^* \rightarrow HH$ process from other sources 
of $HHZ$ production not involving the triple Higgs vertex. In fact, in the first case 
the distribution is peaked just above the $2 M_H$ kinematic threshold and dumped by the 
virtuality of the $H^*$ boson while the background processes exhibit a flatter behaviour
extending up to the upper kinematic limit $\sqrt{s} - M_Z$ (see Figure~\ref{fig:hhz1}).
These characteristics have been tested to discriminate between the signal triple Higgs 
vertex contribution and the background processes in genuine $HHZ$ final states. As the 
phase space increases with increasing $\sqrt{s}$, for a fixed $M_H$ value, this variable
is most effective at larger $\sqrt{s}$ values, i.e. as the dilutiuon effect increases.
A likelihood fit to the total number of events and to the normalised binned distribution
of the $M_{HH}$ distribution has been performed on $HHZ$ events at generator level for
$M_H$ = 120~GeV and $\sqrt{s}$ = 500~GeV and 800~GeV (see Figure~\ref{fig:hhz2}).
Results are given in Table~\ref{tab:hhz}. Work is in progress for improving the 
measurement further by including an additional $e^+e^- \rightarrow HHZ$, where 
$Z \rightarrow \nu\bar{\nu}$ in the final states.

\section{The $HH\nu\bar{\nu}$ process}

\begin{figure}[t]
\begin{tabular}{c c}
\includegraphics[scale=0.4]{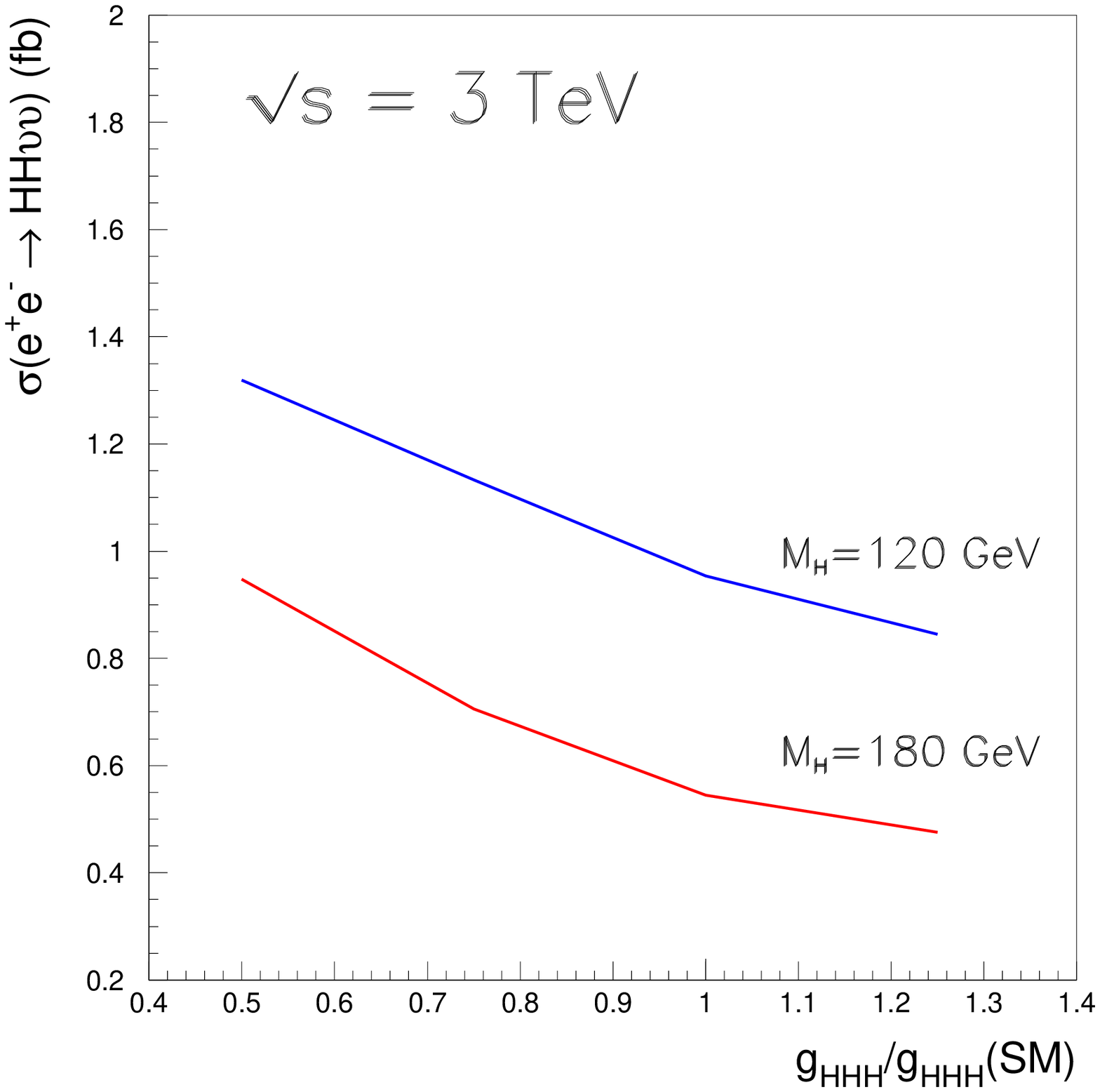} &
\includegraphics[scale=0.4]{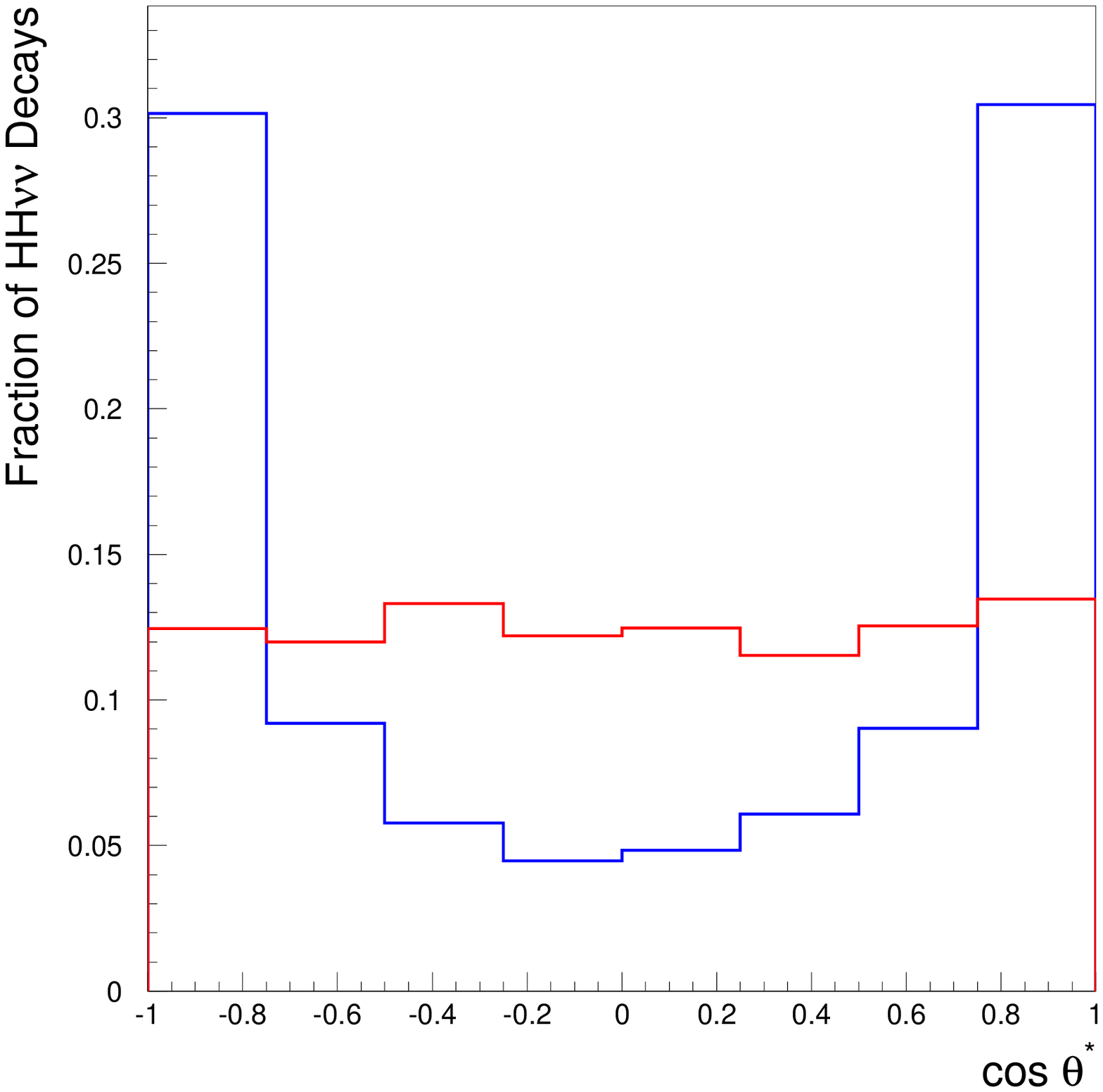} \\
\end{tabular}
\caption{\sl Left: the dependence of the $HH\nu\bar{\nu}$ cross section on the triple 
Higgs coupling, normalised to its SM value, for $\sqrt{s}$ = 3~TeV and two $M_H$ values. 
Right: The $\cos \theta~*$ distribution in $HH\nu\bar{\nu}$ events orginating from 
diagrams containing the triple Higgs vertex (light grey) and other diagrams (dark grey).}
\label{fig:hhnn1}
\end{figure}

\begin{figure}[b]
\includegraphics[scale=0.5]{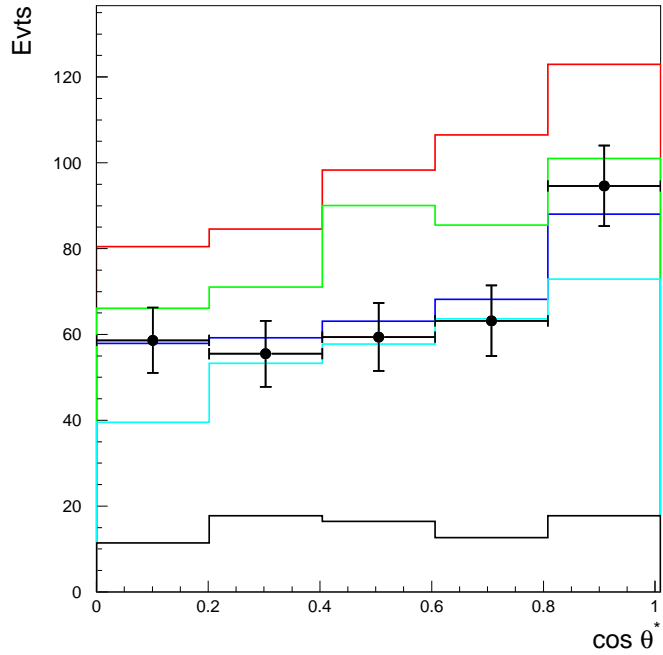}
\caption{The reconstructed $|\cos \theta^*|$ distribution for $HH\nu\bar{\nu}$ events 
obtained for $g_{HHH}/g_{HHH}^{SM}=$1.25,1.0,0.75 and 0.5 with the points with error 
bars showing the expectation for 5~ab$^{-1}$ of SM data at $\sqrt{s}$=3.0~TeV.}
\label{fig:hhnn2}
\end{figure}

The $e^+e^- \rightarrow HH\nu\bar{\nu}$ process exhibits a significant larger 
cross-section, 
compared to $HHZ$, if the collision energy exceeds 1~TeV. In addition, being the 
interference of the signal triple Higgs vertex and the other diagrams, leading to 
$HH\nu\bar{\nu}$, destructive, the total cross section decreases with increasing 
$g_{HHH}$, 
contrary to the case of the $HHZ$ channel. This may represent an important cross-check
of the Higgs self coupling contribution, in the case any deviations would be observed in 
the $HHZ$ cross-section. Since the $HH\nu\bar{\nu}$ production is peaked in the forward 
region, it is important to ensure that an efficient tagging of the 
$H H \rightarrow b \bar b b \bar b$, $W^+W^- W^+W^-$ decay can be achieved. Therefore a 
full reconstruction of $HH\nu\bar{\nu}$ and $ZZ\nu\bar{\nu}$ events has been performed.
The $4~b + E_{miss}$ and $4~W + E_{miss}$ final states have been 
considered for the Higgs boson masses of 120~GeV and 180~GeV.
The sensitivity to the triple Higgs vertex has been enhanced by studying the angle 
$\theta^*$ made by the $H$ boson, boosted back in the $HH$ rest frame, with the $HH$ 
direction. Due to the Higgs boson scalar nature, the $cos \theta^*$ distribution is flat 
for the signal $H^* \rightarrow HH$ process, while it has been found to be forward and 
backward peaked for the $HH\nu\bar{\nu}$ contributions from other 
diagrams~\cite{Boudjema:1996cb}. 
This characteristics is preserved after event reconstruction 
(see Figure~\ref{fig:hhnn1}) and a fit including the measured cross section and the 
normalised shape of the $|\cos \theta^*|$ distribution has been performed, similarly to 
the case of the $HHZ$ channel (see Figure~\ref{fig:hhnn2}). 
Results are given in Table~\ref{tab:hhnn}. Beam polarisation has not been taken into 
account. This may provide the ultimate means for pushing the accuracy on $g_{HHH}$ below 
5\%.

\begin{table}
\caption{Production cross sections for $e^+e^- \rightarrow HHH\nu\bar{\nu}$}
\begin{tabular}{|c|c|c|c|}
\hline
$\sqrt{s}$ & $g_{HHHH}/g_{HHHH}^{SM}$ = 0.9 &  $g_{HHHH}/g_{HHHH}^{SM}$ = 1.0 &
 $g_{HHHH}/g_{HHHH}^{SM}$ = 1.1 \\ 
\hline
~3~TeV & 0.400 & 0.390 & 0.383\\
~5~TeV & 1.385 & 1.357 & 1.321 \\
10~TeV & 4.999 & 4.972 & 4.970 \\ \hline
\end{tabular}
\label{tab:hhhnn}
\end{table}
\begin{table}
\caption{Summary of relative accuracies $\delta g_{HHH}/g_{HHH}$ for $M_H$ = 
120~GeV and $\int L$ = 1~ab$^{-1}$ with the $HHZ$ channel.}
\begin{tabular}{|c|c|c|}
\hline
$\sqrt{s}$ (TeV) & $\sigma_{HHZ}$ Only & $M_{HH}$ Fit \\ \hline
0.5 & $\pm$ 0.23 (stat) & $\pm$ 0.20 (stat) \\
0.8 & $\pm$ 0.35 (stat) & $\pm$ 0.29 (stat) \\ \hline
\end{tabular}
\label{tab:hhz}
\end{table}
\begin{table}
\caption{Summary of relative accuracies $\delta g_{HHH}/g_{HHH}$ for 
$\int L$ = 5~ab$^{-1}$ with the $HH\nu\bar{\nu}$ channel at 3~TeV.}
\begin{tabular}{|c|c|c|}
\hline
$M_H$ (GeV) & $\sigma_{HH\nu\bar{\nu}}$ Only & $|\cos \theta^*|$ Fit \\ \hline
120 & $\pm$ 0.094 (stat) & $\pm$ 0.070 (stat) \\
180 & $\pm$ 0.140 (stat) & $\pm$ 0.080 (stat) \\ \hline
\end{tabular}
\label{tab:hhnn}
\end{table}

\section{Conclusions}

The study of the triple Higgs couplings will provide a crucial test of the Higgs 
mechanism of electro-weak symmetry breaking, by directly accessing the shape of the 
Higgs field potential. $e^+e^-$ linear colliders, where the study of the 
$e^+e^- \rightarrow HHZ$ and $HH\nu\bar{\nu}$ can be performed with good accuracy, 
represent 
a possibly unique opportunity for performing this study. The study of variables sensitive
to the triple Higgs vertex and the availability of high luminosity will allow to test the
Higgs potential structure to an accuracy of about 20\% at a TeV-class collider and to 
8\% and better in multi-TeV $e^+e^-$ collisions. The quartic coupling appears to 
remain unaccessible due to its tiny cross-section and the large dilution effect from 
background diagrams in $HHH\nu\bar{\nu}$.

\begin{acknowledgments}
This study owes much to several collegues with whom we had the pleasure to discuss 
before, during and after the Snowmass workshop. In particular we like to thank 
F.~Boudjema, L.~Maiani, M.~Muhlleitner, K.~Desch and P.~Gay for their suggestions. 
E.B. was supported in part by the CERN-INTAS grant 99-0377 and by the INTAS grant 
00-0679.
\end{acknowledgments}

% Create the reference section using BibTeX:
\bibliography{e3016}

\end{document}